\documentclass[epj]{svjour}

\usepackage{latexsym}
\usepackage{graphics}
\newcommand{\MeV}{\;\mbox{MeV}}
\newcommand{\diag}{\mbox{diag}}
\newcommand{\half}{{\textstyle\frac{1}{2}}}
\newcommand{\twothirds}{{\textstyle\frac{2}{3}}}
\newcommand{\third}{{\textstyle\frac{1}{3}}}

\begin{document}

\title{Phase Structure and Instability Problem in Color
Superconductivity}
\author{Kenji Fukushima}


\institute{RIKEN BNL Research Center, Brookhaven National Laboratory,
 Upton, New York 11973, USA}

\date{Received: date / Revised version: date}

\abstract{
We address the phase structure of color superconducting quark matter
at high quark density.  Under the electric and color neutrality
conditions there appear various phases as a result of the Fermi
surface mismatch among different quark flavors induced by finite
strange quark mass; the color-flavor locked (CFL) phase where quarks
are all energy gapped, the $u$-quark superconducting (uSC) phase where
$u$-quarks are paired with either $d$- or $s$-quarks, the $d$-quark
superconducting (dSC) phase that is the $d$-quark analogue of the uSC
phase, the two-flavor superconducting (2SC) phase where $u$- and
$d$-quarks are paired, and the unpaired quark matter (UQM) that is
normal quark matter without pairing.  Besides these possibilities,
when the Fermi surface mismatch is large enough to surpass the gap
energy, the gapless superconducting phases are expected.  We focus our
discussion on the chromomagnetic instability problem related to the
gapless CFL (gCFL) onset and explore the instability regions on the
phase diagram as a function of the temperature and the quark chemical
potential.  We sketch how to reach stable physical states inside the
instability regions.
} 
\maketitle

\section{Family of color superconducting phases}

  The phase structure of matter composed of quarks and gluons
described by Quantum Chromodynamics (QCD) has been investigated for
many years, and in the high temperature and low baryon (or quark if
deconfined) density region which is accessible in heavy-ion collisions
interesting discoveries have been reported both in theories and in
experiments.  In the high density and low temperature region, on the
other hand, our knowledge is still poor as compared with the rich
physics expected in this region.  Heavy-ion collisions are not
suitable for the purpose to probe dense and cold quark matter, and
such a system could be realized, if any, only in the cores of compact
stellar objects.  The experimental data from the universe is, however,
quite limited and there is no smoking-gun for color superconductivity
so far.  Nevertheless, the theoretical challenge to explore the QCD
phase structure is of great interest on its own.  Also it would be
potentially important in studying the structure and evolution of
neutron stars.

  In order to look into a dense quark system, some of concepts known
in condensed matter physics have been imported into QCD in hope of
analogous phenomena taking place.  In this sense the physics of dense
quark matter is, so to speak, ``condensed matter physics of QCD'' as
articulated clearly in the review~\cite{Rajagopal:2000wf}.
Superconductivity is definitely one of them.  In general the Cooper
instability inevitably occurs wherever there are a sharp Fermi surface
below which particles are degenerated and an attractive interaction
between particles on the Fermi surface.  Even QCD matter is not an
exception and the condensation of quark Cooper pairs leads to color
superconductivity.

  A major difference between ordinary electric superconductivity in
metals and color superconductivity in quark matter arises from the
fact that quarks have three colors and three flavors in addition to
spin, so that quark matter allows for many pairing patterns.  The
color and flavor degrees of freedom make the dense QCD phase structure
so complicated that subtleties still remain veiled.  In this article
we shall argue what has been clarified by now and what should be
solved in the future mainly following the discussions in my recent
papers~\cite{Fukushima:2004zq,Fukushima:2005cm}.

\subsection{Pairing patterns}

  Many theoretical works have revealed that the predominant pairing
pattern is anti-symmetric in spin (spin zero), anti-symmetric in color
(color triplet), and anti-symmetric in flavor (flavor triplet).
Moreover, the color-flavor locking is known to be favored in energy,
so that the color and flavor indices are locked together.  Then there
are three independent diquark condensates or gap
parameters~\cite{Alford:1998mk};
\begin{equation}
 \langle\psi^a_i C\gamma_5\psi^b_j\rangle \sim \Delta_1 \epsilon^{ab1}
  \epsilon_{ij1} + \Delta_2 \epsilon^{abs}\epsilon_{ij2} + \Delta_3
  \epsilon^{ab3}\epsilon_{ij3},
\label{eq:condensate}
\end{equation}
where $(i,j)$ and $(a,b)$ represent the flavor indices $(u,d,s)$ and
the color triplet indices (red,green,blue) respectively.  The charge
conjugation $C$ and the Dirac matrix $\gamma_5$ are required to make
(\ref{eq:condensate}) a Lorenz scalar.  Of course you can 
consider other kind of pairing between quarks which are totally
anti-symmetric under exchange, and even different types of condensates
may coexist.  Actually diquark condensates such as spin-zero pairing
in the color symmetric (color sextet) channel and spin-one pairing
between quarks of the same flavor have been analyzed
quantitatively~\cite{Alford:1999pa,Ruster:2004eg,Iwasaki:1994ij,Schmitt:2004hg}
and known to be much smaller than the predominant condensate.  In this
article we shall simply neglect them.

  Under the pairing ansatz (\ref{eq:condensate}), $\Delta_1$ is a gap
parameter for the Cooper pairing between $d$ and $s$ flavors and green
and blue colors.  That is, $\Delta_1$ is for $bd$-$gs$ and $gd$-$bs$
quarks and $\Delta_2$ and $\Delta_3$ are to be understood likewise;
\begin{center}
\begin{tabular}{rp{5mm}ll}
 $\Delta_1$ && $bd$-$gs$ & $gd$-$bs$ \\
 $\Delta_2$ && $rs$-$bu$ & $bs$-$ru$ \\
 $\Delta_3$ && $gu$-$rd$ & $ru$-$gd$
\end{tabular}
\end{center}

  Each gap parameter is either zero or finite and there are $2^3=8$
combinations accordingly.  Only five of eight phase possibilities as
listed in Table~\ref{tab:phase} are of our interest relevant to the
QCD phase diagram.  When three gap parameters are all nonzero, this
state is called the color-flavor locked (CFL) phase.  When only
$\Delta_1$ is zero, this is the $u$-quark superconducting (uSC) phase
named after the fact that remaining $\Delta_2$ and $\Delta_3$ are gap
parameters for pairing involving $u$-quarks.  The $d$-quark
superconducting (dSC) phase is understood in the same way.  The
two-flavor superconducting (2SC) phase has only $\Delta_3$ which is
nonvanishing.  The question is; where and how they show up on the
phase diagram.  The next section is devoted to this issue.

\begin{table}\begin{center}
\begin{tabular}{lp{1cm}l}
\hline
Gap Parameters && Phase \\
\hline
$\Delta_1\neq0$, $\Delta_2\neq0$, $\Delta_3\neq0$ && CFL\\
$\Delta_1=0$, $\Delta_2\neq0$, $\Delta_3\neq0$ && uSC\\
$\Delta_1\neq0$, $\Delta_2=0$, $\Delta_3\neq0$ && dSC\\
$\Delta_1=\Delta_2=0$, $\Delta_3\neq0$ && 2SC\\
$\Delta_1=\Delta_2=\Delta_3=0$ && UQM\\
\hline\\
\end{tabular}
\end{center}
\caption{A family of color superconducting phases; the color-flavor
locked (CFL) phase, the $u$-quark superconducting (uSC) phase, the
$d$-quark superconducting (dSC) phase, and the two-flavor
superconducting (2SC) phase.  UQM is unpaired quark matter.  The sSC,
2SCsu, and 2SCds phases would not appear in the QCD phase diagram.}
\label{tab:phase}
\end{table}

  It is worth mentioning that these phases can be characterized by
global symmetry breaking patterns. In particular the second-order
phase transitions between the CFL phase and the uSC or dSC phase
belong to the universality class same as an O(2) vector
model~\cite{Fukushima:2005gt}.  In QCD, neither the sSC ($s$-quark
superconducting), 2SCsu (2SC of $s$- and $u$-quarks), nor 2SCds (2SC
of $d$- and $s$-quarks) phase is realized actually because any pairing
containing massive $s$-quarks is disfavored by the Fermi surface
mismatch energy.  (The solution branch of the gap equations belonging
to the 2SCsu phase has been examined in
Refs.~\cite{Alford:2004nf,Abuki:2005ms} and confirmed to cost a larger
energy.)

\subsection{Gapless superconductors}

  The classification of superconducting phases under the pairing ansatz
(\ref{eq:condensate}) is not complete until we will take account of the
gapless superconducting states.  They do not break any new symmetry
and the phase transition to a gapless superconductor exists only at
zero temperature~\cite{Fukushima:2004zq}.  In this sense the gapless
CFL (gCFL) phase for instance which we will closely discuss later is
not a totally new \textit{phase} but can be regarded as a
\textit{variant} of the CFL phase augmented by the presence of gapless
quarks.  Before addressing the CFL problem, we shall see the simplest
example that is actually enough for abstracting the essence.

\begin{figure}\begin{center}
 \resizebox{0.45\textwidth}{!}{%
  \includegraphics{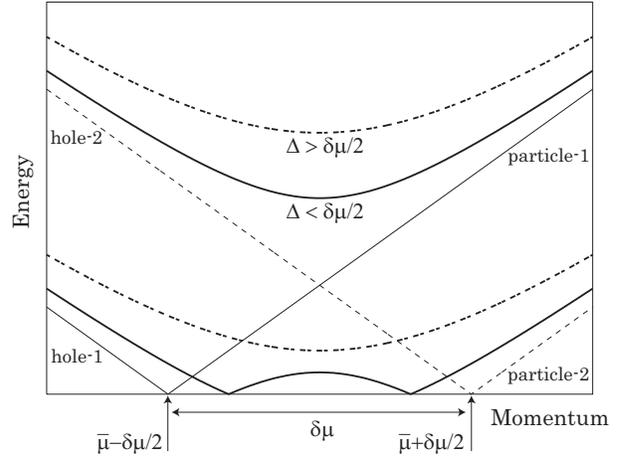}
 }
\end{center}
\caption{The energy dispersion relations near the average Fermi
surface $\bar{\mu}$ for 1 (solid lines) and 2 (dotted lines) species
\textit{without pairing} which have a Fermi surface mismatch
$\delta\mu$.  The dashed curves represent the energy dispersion
relations \textit{with pairing} between 1 and 2 species with the gap
parameter $\Delta>\delta\mu/2$.  When $\Delta<\delta\mu/2$ the
dispersion relations cross zeros as shown by the solid curves and
gapless quarks appear inside the blocking region.}
\label{fig:disp}
\end{figure}

  Let us assume that there are two species of particles, 1 and 2,
which have the Fermi momenta $\bar{\mu}-\delta\mu/2$ and
$\bar{\mu}+\delta\mu/2$ respectively and they form a Cooper pair.  The
energy dispersion relations without pairing are shown by the solid and
dotted lines in Fig.~\ref{fig:disp}.  In the presence of the 1-2 pair
condensate $\Delta$, the level repulsion by the energy $\Delta$
results in the dispersion relations which are smoothly connected
between the hole state of 2 (or 1) and the particle state of 1 (or
2).  In this simple model of superconductivity, therefore, the
quasi-particle energy is expressed as
\begin{equation}
 \epsilon^\pm (p) = \Bigl|\sqrt{(p-\bar{\mu})^2+\Delta^2}
  \pm \half\delta\mu \Bigr|.
\label{eq:disp}
\end{equation}

  As is obvious from Fig.~\ref{fig:disp} as well as from the
expression~(\ref{eq:disp}), the dispersion relation comes to cross
zeros when $|\half\delta\mu|>\Delta$.  The momentum region from one
zero to another zero of the dispersion relation is called the blocking
momentum region because the pairing within this region is hindered by
degenerated particles.  (In the case of Fig.~\ref{fig:disp} particle-2
is degenerated and particle-1 is absent in the blocking region.)
Generally once a superconductor enters the gapless state, the gap
parameter significantly decreases with increasing blocking momentum
region.

  In the language of physics, the condition for the gapless onset
$|\half\delta\mu|>\Delta$ means that a Cooper pair is not stable
energetically.  The pairing energy $2\Delta$ is needed for breaking a
Cooper pair into two particles and at the same time the mismatch
energy $\delta\mu$ is released by doing so.  If the mismatch is more
expensive than the pairing, two particles would no longer form a
Cooper pair.  Roughly speaking, the realization of gapless
superconductivity can be seen as a weak instability disrupting the
Cooper pair only in a limited blocking region but not ruining the
whole superconductivity.

  We will end this subsection with one more remark.  In this article
(and in some literatures) $\delta\mu$ is frequently called the Fermi
surface mismatch.  It should be kept in mind that this mismatch is for
the energy dispersion relations \textit{without} pairing.  In the
presence of pairing, as observed in Fig.~\ref{fig:disp}, the
(approximate) Fermi surface is provided by the average $\bar{\mu}$.
In other words two energy dispersions $\epsilon^\pm(p)$ in
(\ref{eq:disp}) have a common $\bar{\mu}$.

\subsection{Strange quark mass effect and neutrality}

  We have so far illustrated what the gapless superconducting state is
like.  This gapless phase is often called the Sarma phase and was
first demonstrated as a metastable state by Sarma~\cite{Sarma}.  The
Sarma phase has recently attracted much interest since it was pointed
out that it can be stable under the constraint to fix the particle
number~\cite{Liu:2002gi,Gubankova:2003uj}.  Also in the context of QCD
the gapless 2SC (g2SC) phase first~\cite{Shovkovy:2003uu} and the gCFL
phase next~\cite{Alford:2003fq} have been shown to be stable under the
electric and color neutrality conditions, that is; a potential
\textit{maximum} with respect to the gap parameters turns to be a
\textit{minimum} if additional gap parameter dependence through the
neutrality conditions is properly taken into account.  The gapless
superconducting state stabilized in this way, however, turns out to be
unstable after all, which is actually the main subject we shall argue
spending the last half of this article.

  Anyway, postponing the instability problem for a while, let us go on
our discussion on the gCFL phase.  Now we have to understand where
$\delta\mu$ could come from.  There are two distinct contributions
from one origin; one is the direct $M_s$ effect on the energy
dispersion relations and the other is the induced $M_s$ effect through
the neutrality constraints.

  The direct effect is simple.  In the vicinity of the Fermi surface
($p\sim\mu$), the quark energy dispersion relation can be well
approximated as
\begin{equation}
 \sqrt{p^2+M_s^2}-\mu \simeq p-\Bigl(\mu-\frac{M_s^2}{2\mu}\Bigr).
\label{eq:approx}
\end{equation}
Thus the strange quark mass effect can be incorporated by a chemical
potential shift proportional to $M_s^2/\mu$.  It is actually the case
that the physics should change as a function of $M_s^2/\mu$ alone.

  An explanation of the induced $M_s$ effect needs some knowledge on
the neutrality conditions.  As far as a chunk of quark matter, like
the cores of neutron stars which are of kilometer size, is concerned,
neutrality with respect to charge associated with any gauge field must
be required~\cite{Iida:2000ha,Alford:2002kj,Steiner:2002gx}.  Besides,
in the case of color charge in QCD, bulk quark matter must be a color
singlet as a whole system.  This is a more stringent condition than
neutrality.  The singletness is in principle to be implemented by an
appropriate projection operator onto neutral states.  There are many
quarks in a macroscopic system and there are many thermodynamically
equivalent states.  If the number of singlet states grows in the large
volume limit in the same way as the number of neutral states, the
singletness condition can be simply ignored in thermodynamic
properties.  This is in fact the case in color superconducting quark
matter~\cite{Amore:2001uf}.  Thus only the neutrality conditions for
electric and color charges suffice for our purpose to explore the
phase structure.

  In the grand canonical ensemble, the number density of particles is
specified by a chemical potential.  We should introduce one chemical
potential $\mu_e$ for the electromagnetic charge and eight chemical
potentials $\mu_\alpha$ for the color charges which are not
gauge-invariant in non-abelian gauge theories.  For the purpose of
imposing neutrality it is enough to constrain only two eigenvalues of
the color charge since zero charge remains to be zero charge if it
is rotated by any gauge transformations.  We thus only have to solve
the color neutrality conditions with respect to two chemical
potentials $\mu_3$ and $\mu_8$~\cite{Alford:2002kj}.

  Quark matter with three flavors would be automatically electric and
color neutral if $M_s$ is zero, meaning that $\mu_e$, $\mu_3$, and
$\mu_8$ should be of order $M_s^2/\mu$ just like the direct $M_s$
effect.  

  In summary, the direct and induced $M_s$ effects are concisely
expressed in a form of the effective chemical potentials for
respective quarks with color $a$ and flavor $i$ as
\begin{eqnarray}
 \mu_{a i} &=& \mu - \mu_e \bigl(Q\bigr)_{ii}
  + \mu_3 \bigl(T_3\bigr)_{aa} \nonumber\\
 && \qquad + \mu_8 \bigl(\twothirds T_8\bigr)_{aa}
  - \bigl(\vec{M}^2\bigr)_{ii}/2\mu,
\label{eq:chemical}
\end{eqnarray}
where $Q=\diag(\twothirds,-\third,-\third)$ and
$\vec{M}=\diag(0,0,M_s)$ in flavor ($u$,$d$,$s$) space and
$T_3=\diag(\half,-\half,0)$ and
$\frac{2}{\sqrt{3}}T_8=\diag(\third,\third,-\twothirds)$ in color
(red,green,blue) space.  From (\ref{eq:chemical}) the Fermi surface
mismatch for all the pairings is easily inferred;  for the $bd$-$gs$
pairing for instance the mismatch is
\begin{equation}
 \delta\mu_{bd\mathrm{-}gs}=\mu_{bd}-\mu_{gs}
  ={\textstyle\frac{1}{2}}\mu_3-\mu_8+{\textstyle\frac{M_s^2}{2\mu}}.
\end{equation}
In the CFL phase at zero temperature a model-independent
argument~\cite{Alford:2002kj} yields $\mu_e=\mu_3=0$ and
$\mu_8=-M_s^2/2\mu$, and together with the above expression, the
gapless onset condition (where gapless quarks begin appearing) is
\begin{equation}
 \delta\mu_{bd\mathrm{-}gs}=M_s^2/\mu>2\Delta_1.
\label{eq:onset}
\end{equation}
Numerical calculations in a model
study~\cite{Fukushima:2004zq,Alford:2003fq} have confirmed this onset
condition being a good estimate.

\section{Phase diagram of dense quark matter}

\begin{figure*}\begin{center}
 \resizebox{0.8\textwidth}{!}{%
  \includegraphics{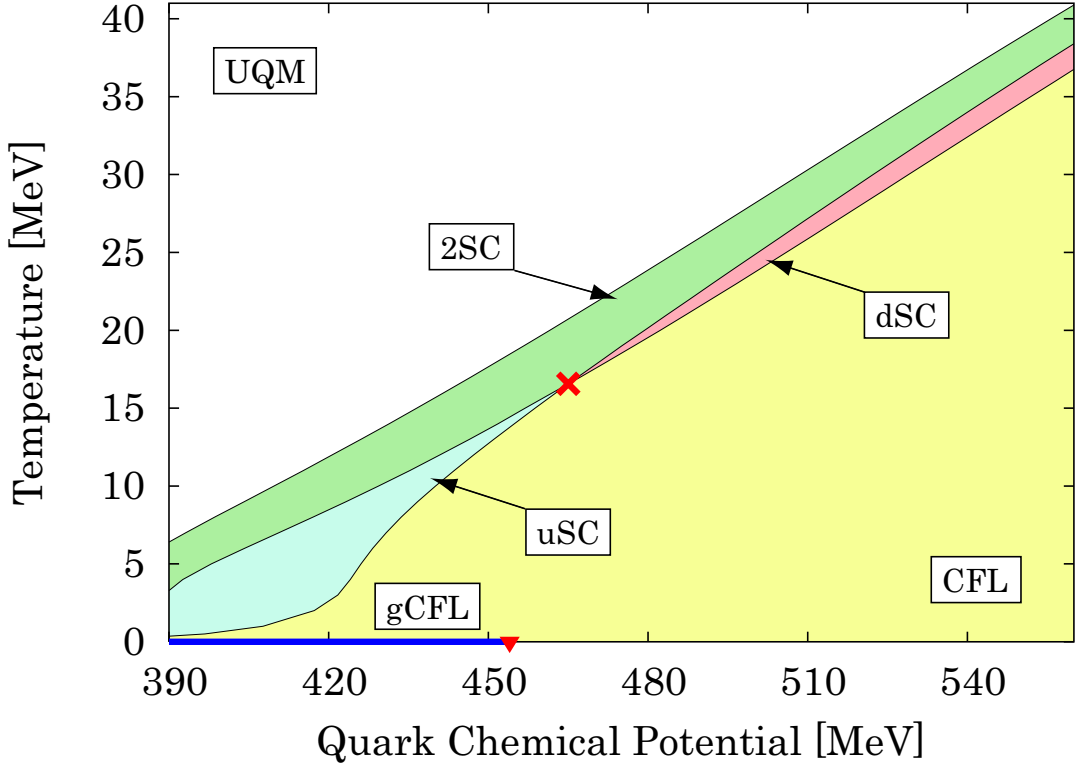}
 }
\caption{Phase diagram of dense quark matter obtained in the NJL model
with the parameters chosen to give $\Delta=40\MeV$ at $\mu=500\MeV$
and $T=M_s=0$.  The strange quark mass $M_s$ is fixed at $150\MeV$
under the assumption that the first-order chiral phase transition
occurred at some chemical potential below $390\MeV$ and
$\langle\bar{s}s\rangle\propto M_s$ above that would not induce large
corrections to $M_s$.  The doubly critical point and the gCFL onset
are marked by a cross and a triangle respectively.}
\label{fig:phase_dg_base}
\end{center}\end{figure*}

  We will present the phase diagram first and then look closely at
each phase in turn.  Figure~\ref{fig:phase_dg_base} is the phase
diagram obtained in the Nambu--Jona-Lasinio (NJL) model.  The model
parameters are chosen as to yield $\Delta=40\MeV$ at $\mu=500\MeV$ and
$M_s=T=0$. The strange quark mass is fixed at $M_s=150\MeV$ which is
the lowest estimate for the $M_s$ effect in the intermediate density
region.  In Ref.~\cite{Abuki:2004zk} the authors solved the gap
equations for the chiral condensates as well as the gap parameters and
the chemical potentials.  In the case of weak coupling, $\Delta$ is
smaller and thus the critical value of the strange quark mass
(or chemical potential) on the gCFL onset is smaller (or larger), and
the gCFL phase is not substantially affected by the chiral dynamics
then.

  Even if the chiral condensate is not dealt with dynamically, a weird
structure including a first-order phase boundary emerges on the phase
diagram for large $M_s^2/\mu$.  As discussed in
Ref.~\cite{Fukushima:2004zq}, however, the complicated structure seen
at large $M_s^2/\mu$ is model and parameter dependent.  Unfortunately
there is no guiding principle to avoid unphysical artifacts in a
certain parameter choice and one has to try and show all the cases for
conclusive analyses~\cite{Abuki:2004zk}.  In this article we dare not
to touch that subtle issue and actually this is the reason why we
present the phase diagram only for $\mu>390\MeV$; a first-order phase
boundary between the gCFL phase and the UQM is found at
$\mu\simeq388\MeV$.

  The phase structure depicted in Fig.~\ref{fig:phase_dg_base} is
robust at least in a sense that the gross features would not be
amended by assumptions we have made to draw this figure.  Besides the
chiral dynamics the important physics we have dropped is the
$K^0$-condensation that could delay the gapless
onset~\cite{Kryjevski:2004jw} and the 't~Hooft interaction term that
embodies the instanton induced $\mathrm{U_A}(1)$ breaking interaction
in the NJL model Lagrangian.  These two missing effects are actually
closely linked.  They could change the quantitative details of the
phase diagram but no qualitative essence would be altered.  We would
not refer to the mixed phase possibility here, for it has already been
settled~\cite{Alford:2004nf}.

  We can see that in the lower density region near the gCFL phase the
uSC region opens at finite temperature while the dSC phase is relevant
at higher density.  We can understand this behavior in a
model-independent way and the existence of the ``doubly critical
point'' is concluded~\cite{Fukushima:2004zq}.  In the subsequent
subsections we shall discuss the important nature of the gCFL phase
and how the uSC and dSC phases come out from the Fermi surface
mismatch, and finally comment upon the doubly critical point.

\subsection{gCFL phase}

  The gapless onset is located at $\mu\simeq455\MeV$ where
$\Delta_1=24.9\MeV$ and $\mu_8=-24.7\MeV$ and one can readily confirm
that $\mu_8\simeq-M_s^2/2\mu$ and $M_s^2/\mu\simeq 2\Delta_1$ as
indicated by the condition~(\ref{eq:onset}).  The gCFL phase has a
definite meaning only at zero temperature because there is no clear
distinction between gapless quarks and thermally excited quarks at
finite temperature.

  The onset condition~(\ref{eq:onset}) is only for $bd$-$gs$ quarks
with $\Delta_1$ but gapless quarks can be present in another quark
sector as well.  In fact the energy dispersion relations of $rs$-$bu$
quarks turn out to be identical to those of $bd$-$gs$ quarks as long
as the system remains to be in the CFL (not gCFL) phase.  Thus
$rs$-$bu$ quarks become gapless at $M_s^2/\mu\simeq 2\Delta_2$ where
$\Delta_2=\Delta_1$ holds in the CFL phase.  For larger $M_s$ or
smaller $\mu$, the blocking momentum region in the $bd$-$gs$ sector
becomes wider, while the blocking region in the $rs$-$bu$ sector is
severely constrained by the neutrality conditions so that it must
remain tiny (see
Refs.~\cite{Fukushima:2004zq,Fukushima:2005cm,Alford:2003fq} for
details).  As a result the $rs$-$bu$ dispersion
relations are kept to be almost quadratic in the entire gCFL region.
Therefore only $\Delta_1$ in the $bd$-$gs$ quark sector significantly
drops in the gCFL region due to the spreading blocking region with
increasing $M_s^2/\mu$.

  Because this is an essential point in considering the instability
problem later, we shall reiterate here;
\vspace{2mm}

\noindent
 $bd$-$gs$ pairing with $\Delta_1$ ---
  gapless quarks appear at $M_s^2/\mu=2\Delta_1$; the blocking region
  increases for larger $M_s^2/\mu$.
\vspace{2mm}

\noindent
 $rs$-$bu$ pairing with $\Delta_2$ ---
  gapless quarks appear at $M_s^2/\mu=2\Delta_2$; the dispersion
  relations are kept to be almost quadratic in the entire gapless
  phase.
\vspace{2mm}

\noindent
 $gu$-$rd$ pairing with $\Delta_3$ ---
  gapped quarks only.
\vspace{2mm}

\noindent
 $ru$-$gd$-$bs$ pairing with all $\Delta$'s ---
  gapped quarks only.

\subsection{uSC phase}

  The uSC phase results from the presence of the gCFL phase at zero
temperature.  In the gCFL phase, as explained in the last section,
$\Delta_1$ is significantly reduced than $\Delta_2$ and $\Delta_3$.
The pairing between $u$-$d$ quarks is not reduced but $\Delta_3$ is
enhanced in the gCFL side as $\Delta_1$ and $\Delta_2$
decreases~\cite{Fukushima:2004zq}.  (Note that $\Delta_3$ evaluated in
the 2SC state is in general larger than $\Delta_3$ in the CFL state
for the same diquark interaction.)  In this way the ordering
$\Delta_1<\Delta_2<\Delta_3$ is realized.

  It is a well-known fact that the critical temperature is of order of
the gap parameter at zero temperature.  We can anticipate that
$\Delta_1$ would melt first at finite temperature right above the gCFL
region on the phase diagram.  This means that the uSC phase where only
$\Delta_1$ is zero is expected when gCFL matter is heated.
Figure~\ref{fig:phase_dg} clearly shows that the phase structure is
certainly as anticipated.

\subsection{dSC phase}

  It is not the Fermi surface mismatch but the average Fermi momentum
that is more relevant away from the gCFL region (see the
expression~(\ref{eq:disp}) or Fig.~\ref{fig:disp}).  At zero
temperature the average Fermi momenta are common in all the quark
sectors, that leads to equality in the number of nine (three colors
and three flavors) quarks, and thus the electric and color neutrality
is enforced~\cite{Rajagopal:2000ff}.

  The enforced neutrality at zero temperature is broken at small
temperatures of a few MeV~\cite{Fukushima:2004zq}.  At higher
temperature, especially in the vicinity of the critical temperatures,
it is a good approximation to estimate $\mu$'s in the normal phase,
i.e., $\mu_e=-M_s^2/4\mu$ and $\mu_3=\mu_8=0$.  Then the ordering of
the Fermi momenta, $\mu_s<\mu_u<\mu_d$ is concluded, which can be
understood in an intuitive way; the number of $s$-quarks is suppressed
by $M_s$ and so $d$-quarks should be more abundant than $s$-quarks to
maintain electric neutrality.  From this, the average Fermi momenta
should obey the following ordering;
$\bar{\mu}_{su}<\bar{\mu}_{ds}<\bar{\mu}_{ud}$.

  The gap equation to determine $\Delta$'s contains the momentum
integration around the Fermi surface which effectively picks up the
density of states at the Fermi momentum.  The larger the density of
states is, the greater the gap parameter becomes.  In this way the
ordering $\Delta_2<\Delta_1<\Delta_3$ is realized from the average
Fermi momenta ordering, which follows the presence of the dSC phase
accordingly, as first discussed in Ref.~\cite{Iida:2003cc}.

\subsection{Doubly critical point}

  The phase boundary on which $\Delta_1$ goes to zero crosses the
phase boundary on which $\Delta_2$ goes to zero at the ``doubly
critical point'' where two phase transitions with respect to
$\Delta_1$ and $\Delta_2$ take place simultaneously.  Since the
existence of the uSC and dSC phases are robust and model-independent,
so is the doubly critical point.

  We would shortly comment upon a puzzling question concerning the
doubly critical point.  The question is the following; what are the
effects of gauge field fluctuations on the doubly critical point?

  In weak coupling the gauge field fluctuations bring about an induced
first-order phase transition and the critical temperature is
shifted~\cite{Matsuura:2003md,Giannakis:2004xt}.  Between the CFL
phase and the uSC or dSC phase, no manifest effects would be expected
because eight gluons are all massive in both phases.  Therefore we can
conclude that the phase transitions from the CFL phase toward either
the uSC or dSC phase is surely of second order belonging to the same
universality class as an O(2) vector model.
  
  The gauge field fluctuations play an important role, on the other
hand, between the 2SC phase and the uSC or dSC phase and the phase
transition is forced to be of first order.

  The doubly critical point is the point at which two phase
transitions meet and three gluons become massless around it since it
faces the 2SC phase.  This suggests that the phase transition must be
of first order and the critical temperature should be shifted at the
doubly critical point.  But how is it possible if the phase boundaries
facing the CFL phase are not shifted at all?  Or, are they shifted
near the doubly critical point by nearly massless gluons?  To answer
this question, further clarification on the treatment of the gauge
field fluctuations is waited.

\section{Chromomagnetic instability}

\begin{figure*}\begin{center}
 \resizebox{0.8\textwidth}{!}{%
  \includegraphics{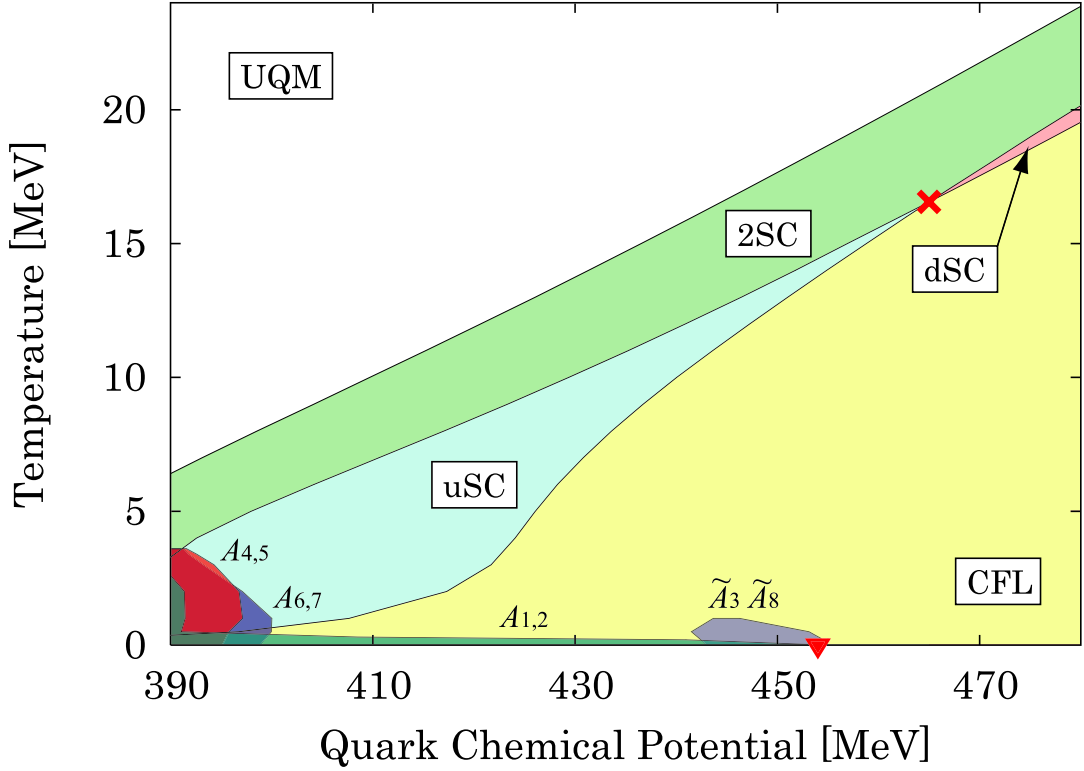}
 }
\caption{A magnified drawing of the phase diagram around the gCFL and
uSC phases with the instability regions overlaid.}
\label{fig:phase_dg}
\end{center}\end{figure*}

  The gapless superconductors could be stabilized by the neutrality
conditions, at least, in the variational space spanned by gap
parameters and chemical potentials, but after all, they have turned
out to be unstable in a wider space including the gauge fields.  The
chromomagnetic instability signifies that the Meissner screening mass
squared becomes negative, that is, the Meissner mass is imaginary.
Before addressing the chromomagnetic instability problem first pointed
out by Huang and Shovkovy~\cite{Huang:2004bg} in the case of the g2SC
phase and analyzed in
Refs.~\cite{Fukushima:2005cm,Casalbuoni:2004tb,Alford:2005qw} for the
gCFL phase, we shall make a brief overview on the Debye and Meissner
screening masses in the CFL phase, which would be useful to make a
deeper insight into the instability problem.  Our results are
summarized in Fig.~\ref{fig:phase_dg} and the goal of this section is
to explain what is going on inside the instability regions presented
by shadowed regions in this figure.

\subsection{Debye and Meissner screening masses}

  The Debye and Meissner screening masses are the screening masses for
the longitudinal and transverse gauge fields respectively, which are
defined by
\begin{eqnarray}
 m_{D,\alpha\beta}^2 &=& -\lim_{q\to0}\Pi_{\alpha\beta}^{00}
  (\omega=0,\vec{q}),\\
 m_{M,\alpha\beta}^2 &=& {\textstyle\frac{1}{2}}\lim_{q\to0}
  (\delta_{ij}-\widehat{q}_i\widehat{q}_j)\Pi_{\alpha\beta}^{ij}
  (\omega=0,\vec{q}),
\end{eqnarray}
where $\widehat{q}_i=q_i/|\vec{q}|$ and $\Pi_{\alpha\beta}^{\mu\nu}$
is the polarization tensor for the gauge fields $A_\alpha^\mu$ with
the color and Lorenz indices denoted by $\alpha$ and $\mu$.  At high
enough density the polarization tensor is dominated by the quark-loop
contributions alone and so we shall neglect the gluon and ghost loops
that would not depend on $\mu$ at the one-loop level.

  Let us take one example to elaborate the quark-loop contributions
for each gluon.  The $A_1$ and $A_2$ gluons with the color indices
labeled according to the Gell-Mann matrices in color space couple red
quarks to green quarks and vice versa, while the flavor is not
changed at any gluon vertices.  For example, a diagram graphically
shown as
\begin{center}
 \resizebox{0.3\textwidth}{!}{%
  \includegraphics{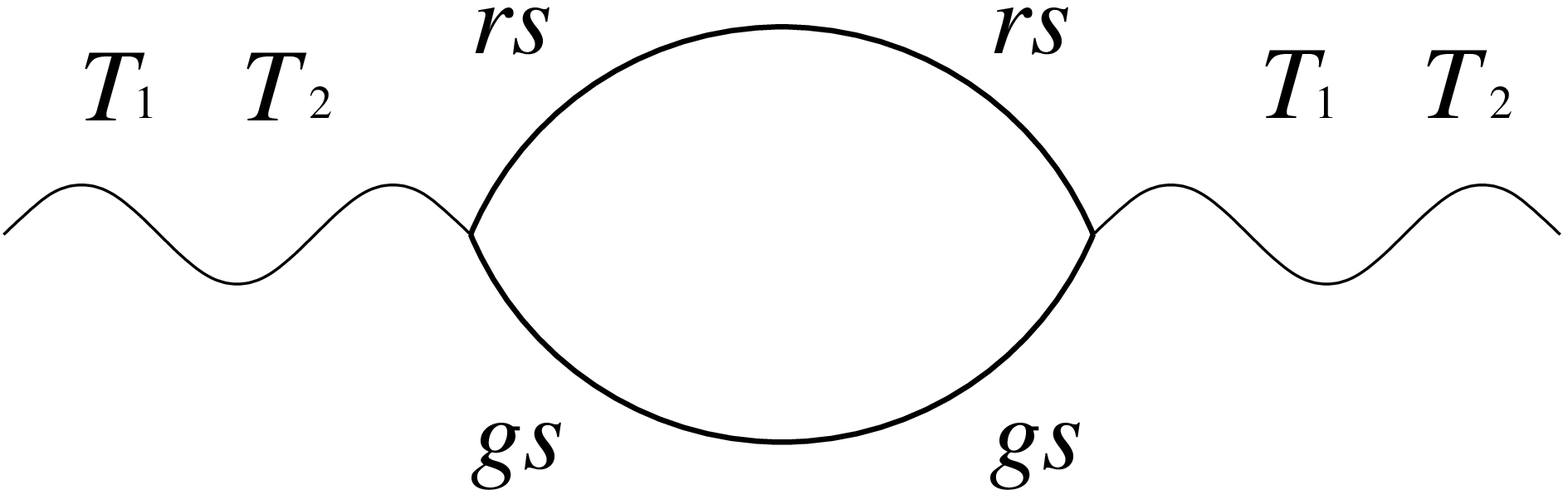}} (Example 1)
\end{center}
can contribute to $\Pi_{11}^{\mu\nu}$ and $\Pi_{22}^{\mu\nu}$.  Solid
lines represent the quark propagator for $rs$ and $gs$ quarks in this
specific case.  From this kind of diagrammatic deliberation, it is
easy to confirm that there is no mixing between gluons except
$A_3$-$A_8$ mixing.  Moreover, some algebra can readily tell us that
the screening masses are degenerate for the gluon pairs ($A_1$,$A_2$),
($A_4$,$A_5$), and ($A_6$,$A_7$).  In this article, like in
Ref.~\cite{Fukushima:2005cm}, we shall denote $A_{1,2}$, $A_{4,5}$,
and $A_{6,7}$ to mean either of the degenerated pairs.

\subsection{Unstable channels and mixing with photon}

  If there is something funny in connection with gapless quarks in the
quark-loop polarization like Example~1, it is expected to happen when
both two quark propagators involve gapless energy dispersion
relations.  At the gapless onset, in fact, the quark-loop polarization
consisting only of $bd$, $gs$, $rs$, and $bu$ quarks is divergent,
leading to the divergent contributions to the Debye and Meissner
masses for $A_{1,2}$, $A_3$, and $A_8$ gluons.  You can easily check
that the polarization diagrams relevant to the $A_{4,5}$ and $A_{6,7}$
gluon channels must have at least one of $gu$, $rd$, $ru$, $bd$, and
$bs$ quarks which are gapped.

  It should be noted here about mixing between the $A_\gamma$ photon
and the $A_3$ and $A_8$ gluons.  In the symmetric CFL phase with
$M_s=0$, there is no mixing with respect to $A_3$, but in the present
case with $M_s\neq0$ and thus $\mu_e\neq0$ in the gCFL phase, there is
$A_3$-mixing as well due to the isospin symmetry breaking.  The mass
squared matrix for $A_\gamma$, $A_3$, and $A_8$ is a $3\times3$ matrix
with nonvanishing off-diagonal components generating mixing.  We will
denote the eigenmodes of the mass squared matrix by
$\widetilde{A}_\gamma$, $\widetilde{A}_3$, and $\widetilde{A}_8$.  The
rotated photon represented by the $\widetilde{A}_\gamma$ field is
massless all the way because the CFL and gCFL phases preserve a
rotated electromagnetic U(1) symmetry.  Once mixing occurs among
$A_\gamma$, $A_3$, and $A_8$, it is simply a matter of convention
which eigenmode should be identified as $\widetilde{A}_3$ or
$\widetilde{A}_8$.  By this reason, though Figure~\ref{fig:phase_dg}
has the instability region with the label $\widetilde{A}_3$ and
$\widetilde{A}_8$, it does not mean that both of two eigenmodes suffer
from the instability, but the fact is actually that only one of them
does.

\subsection{Divergences at the gCFL onset}

The divergences in the Debye and Meissner screening masses at the
gapless onset derive from the density of states which is divergent
when the energy dispersion relations take a quadratic form;
$\epsilon(p)\sim(p-\bar{\mu})^2/2\Delta$.  That is, the density of
states $n(p)$ is given by
\begin{equation}
 n(p) = 4\pi p^2\biggl[\frac{d\epsilon(p)}{dp}\biggr]^{-1}
\end{equation}
and obviously $n(p)\to\infty$ when the slope of $\epsilon(p)$ is zero
at $p=\bar{\mu}$.

  The next question is whether the divergence is positive or negative
in the Debye and Meissner masses.  A naive intuition would be that
both are positive, for in the CFL phase it is well established that
$m_M^2=\frac{1}{3}m_D^2$ should
hold~\cite{Fukushima:2005gt,Son:1999cm,Rischke:2000ra} and one might
well consider that they are correlated in a similar way even in the
presence of $M_s\neq0$.  This expectation is partly true but in an
unexpected manner as we will shortly see below.

  To find the relations between the Debye and Meissner masses, it is
convenient to split the quark propagator into four distinct parts.  If
the quark mass effect is approximated as (\ref{eq:approx}) near the
Fermi surface, the energy projection operator divides the propagator
into the \textit{particle} part that is a function of $p-\bar{\mu}$
and the \textit{antiparticle} part that is a function of
$p+\bar{\mu}$. Moreover, in the Nambu-Gor'kov formalism, the quark
propagator is a $2\times2$ matrix and its \textit{diagonal} component
is a normal propagation of particles, and its \textit{off-diagonal}
component is an abnormal propagation mediated by diquark
condensates. There are thus four distinct combinations;
diagonal--particle, diagonal--antiparticle, off-diagonal--particle,
and off-diagonal--antiparticle.

  The diagram shown in Example~1 is one example consisting of only the
diagonal propagators.  We can construct another example in the same
gluon channel composed of only the off-diagonal propagators as
follows;
\begin{center}
 \resizebox{0.3\textwidth}{!}{%
  \includegraphics{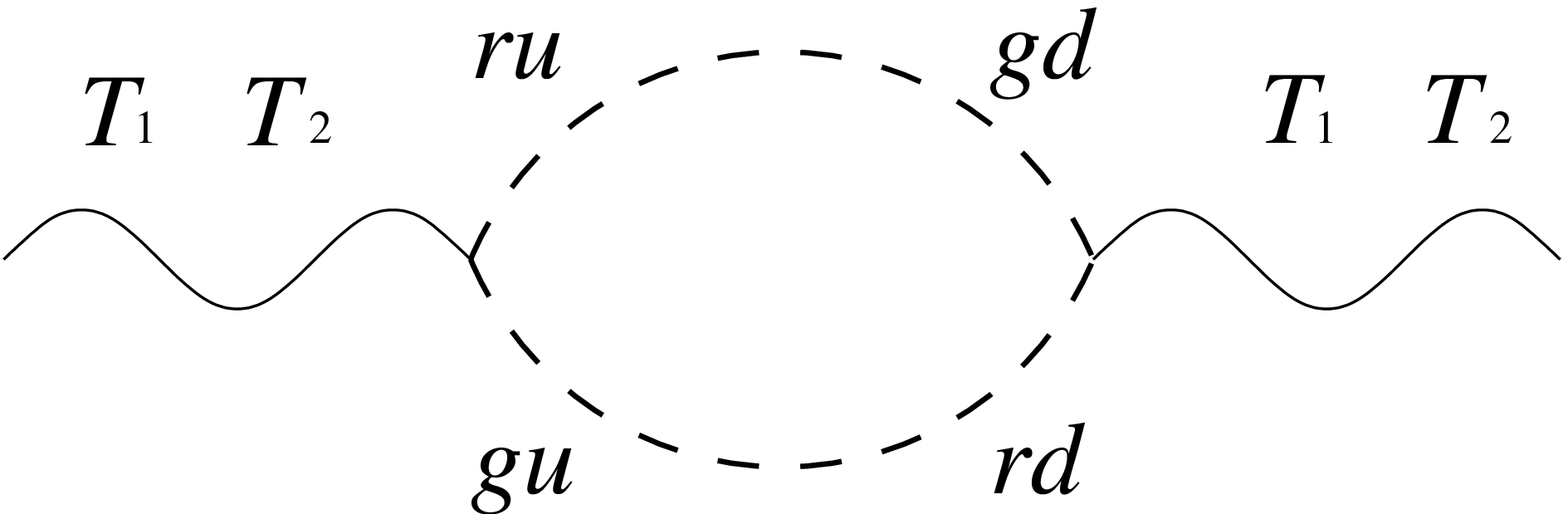}} (Example 2)
\end{center}
where the upper propagator is the off-diagonal component proportional
to $\Delta$'s connecting $ru$ and $gd$ quarks, and the lower one is
the off-diagonal component proportional to $\Delta_3$ connecting $gu$
and $rd$ quarks.

  It is important to note that the possible singular behavior
originating from gapless quarks can reside only in the particle parts
because antiparticles would never be gapless.  The Debye mass squared
receives the contributions only from particles which can be separated
into the diagonal part, $[m_D^2]_{\mathrm{diag}}$, and the
off-diagonal part, $[m_D^2]_{\mathrm{off}}$.  The Meissner mass, on
the other hand, has all of the particle-particle,
particle-antiparticle, and antiparticle-antiparticle contributions.
In Ref.~\cite{Fukushima:2005cm} interesting relations have been found;
\begin{eqnarray}
 \bigl[m_M^2\bigr]_{\mathrm{diag(pp)}} &=& -{\textstyle\frac{1}{3}}
  \bigl[m_D^2\bigr]_{\mathrm{diag}}\;,
\label{eq:diag} \\
 \bigl[m_M^2\bigr]_{\mathrm{off(pp)}} &=& {\textstyle\frac{1}{3}}
  \bigl[m_D^2\bigr]_{\mathrm{off}}\;,
\label{eq:off}
\end{eqnarray}
where $[m_M^2]_{\mathrm{diag(pp)}}$ is a part of the Meissner mass
squared coming from the diagrams with two diagonal particle
propagators (like Example~1) and $[m_M^2]_{\mathrm{off(pp)}}$ with two
off-diagonal particle propagators (like Example~2).  The
relation~(\ref{eq:diag}) might be strange at a glance, for we already
know that $m_M^2=\third m_D^2$ in the CFL phase at $M_s=0$.  Then,
how can we retrieve the relation $m_M^2=\third m_D^2$?  Actually, as
long as the system stays in the CFL side, the diagonal
particle-antiparticle contribution $[m_M^2]_{\mathrm{diag(pa)}}$ is
\textit{twice larger} than $[m_M^2]_{\mathrm{diag(pp)}}$ and changes
the overall sign in the relation between the Debye and Meissner masses.
The other contributions, $[m_M^2]_{\mathrm{diag(aa)}}$,
$[m_M^2]_{\mathrm{off(pa)}}$, and $[m_M^2]_{\mathrm{off(aa)}}$ are all
negligibly small.

  It might be surprising that these relations (\ref{eq:diag}) and
(\ref{eq:off}) are satisfied as they are even in the gCFL phase!  The
diagonal and off-diagonal parts both can generate divergent
contributions at the gCFL onset.  As far as the QCD problem is
concerned, the diagonal contributions are always larger than the
off-diagonal ones, leading to the Debye mass squared diverging
\textit{positively} and Meissner mass squared diverging
\textit{negatively} at the gCFL onset.  The negative sign in
(\ref{eq:diag}) cannot be compensated in the gCFL phase since, unlike
in the CFL phase, $[m_M^2]_{\mathrm{diag(pa)}}$ is no longer a match
for $[m_M^2]_{\mathrm{diag(pp)}}$.

  To put it in another way, the key relation (\ref{eq:diag}) can be
stated as follows:  In the diagonal part, we can say that the
particle-particle loops tend to induce \textit{paramagnetism}, while
\textit{diamagnetism} stems from the particle-antiparticle loops.
Usually in the superconducting phase, the diamagnetic tendency is
greater enough to exhibit the Meissner effect.  In gapless
superconductors, however, antiparticles are never gapless and only the
particle-particle loops are abnormally enhanced due to the presence of
gapless quarks.  As a result of that, the diamagnetism gives way to
the chromomagnetic instability.

\subsection{Away from the gCFL onset}

  Figure~\ref{fig:phase_dg} shows the instability regions for
respective gluons.  The difference between the $A_{1,2}$ behavior and
the $\widetilde{A}_3$-$\widetilde{A}_8$ behavior can be understood
from the difference between the $bd$-$gs$ and $rs$-$bu$ quark
dispersion relations.  It is only the diagram shown in Example~1 that
causes singular behavior in $A_{1,2}$ around the gCFL onset.  This
diagram has two quark propagators; one of \textit{gapless} $gs$-quarks
and the other of \textit{quadratic} $rs$-quarks.  Since $rs$ quarks
are kept to be almost quadratic in the entire gCFL region, as we have
explained before, the $A_{1,2}$-instability extends over the whole
gCFL region at small temperatures.  The instability would not persist
into regions at higher temperature because the quadratic dispersion
relations are easily affected by thermal excitations.

  In contrast, the same species of quarks constitute the
$\widetilde{A}_3$-$\widetilde{A}_8$ instability region.  The
instability caused by two quadratic quark propagations lies in the
entire gCFL region like the $A_{1,2}$-instability but only at tiny
temperatures of order eV.  Instability boundaries at such low
temperatures are not visible actually on the phase diagram.  The
instability caused by two gapless quark propagations is localized near
the gCFL onset and spreads toward high temperatures than the
$A_{1,2}$-instability because it has nothing to do with quadratic
quarks.

  Although we would not go further into details, the Meissner
screening masses evaluated in the g2SC phase~\cite{Huang:2004bg}
indicate that the $\widetilde{A}_8$-instability occurs starting at
the g2SC onset and there also arises the instability for $A_{4,5,6,7}$
gluons whose boundary is found not at the g2SC onset but inside the
2SC region.  In our results the $A_{4,5}$ and $A_{6,7}$ instability
regions are located at large Fermi surface mismatch and enter the uSC
phase at higher temperature and then the 2SC phase farther.  We
conjecture that these instability regions for $A_{4,5}$ and $A_{6,7}$
would be linked to the instability found in the 2SC phase.  The nature
of the instability with respect to $A_{4,5}$ and $A_{6,7}$ is
presumably different from the instability near the gCFL onset that we
have seen in great details.  In the aim of disclosing the QCD phase
diagram in the intermediate density region, in particular, the
$A_{4,5}$ and $A_{6,7}$ instability deserves further investigation.

\section{Speculations}

  This final section is devoted to sketching some speculations on how
to reach the stable states inside the instability regions on the phase
diagram, which has been barely succeeded so far.

  There are already some attempts to interpret and resolve the
chromomagnetic instability
problem~\cite{Giannakis:2004pf,Huang:2005pv,Hong:2005jv,Alford:2005kj}.
The most important among them is, in my opinion, the observation
pointed out by Giannakis and Ren in Ref.~\cite{Giannakis:2004pf} that
the instability with respect to gluons can be interpreted as the
instability toward a plane-wave crystalline superconducting phase.

  The most straightforward interpretation of the chromomagnetic
instability is, of course, spontaneous generation of the expectation
value for the transverse gluons.  The gauge field itself is, however,
not a physical quantity depending on the gauge choice.  Actually, what
Giannakis and Ren realized in Ref.~\cite{Giannakis:2004pf} is that the
spontaneously generated gauge fields can be absorbed in the phase of
the gap parameters by means of the gauge transformation.  It should be
noted that whether the gauge invariance is maintained or not does not
matter.  The gauge transformation in this manipulation simply means
the change of variables.

  The crystalline phase had been already studied~\cite{crystalline}
before the chromomagnetic instability was discovered.  It is called
the ``crystalline'' phase because the gap parameter takes a form of
\begin{equation}
 \Delta(x) = |\Delta|e^{i\vec{q}\cdot\vec{x}},
\label{eq:crystalline}
\end{equation}
which breaks the translational and rotational invariance.  The
essential point is that there is another quark basis where the phase
factor of (\ref{eq:crystalline}) vanishes at the price of the vector
potential arising in the effective action.  Therefore, taking care of
a vector potential turns out to be equivalent with dealing with the
gap parameter with a phase factor corresponding to the given vector
potential.

  Let us rephrase the above mentioned idea in a slightly different
way.  Supposing we performed the derivative expansion of the
thermodynamic potential $\Omega_A[\Delta(x)]$ which is now
\textit{gauged} with gluons, then we have in general,
\begin{eqnarray}
 &&\Omega_A[\Delta(x)] \simeq \Omega_0[\Delta] \nonumber\\
 && - \kappa^{ab}[\Delta] \mathrm{Tr}\bigl[(\partial_i
  \!+\!iA_i^\ast)\Delta^\ast(x)\bigr]^a \bigl[(\partial^i
  \!-\!iA^i)\Delta(x)\bigr]^b \,,
\label{eq:expansion}
\end{eqnarray}
where $a$ and $b$ represent the color triplet indices and
$A^i=A^i_\alpha T^\alpha$ with the color adjoint index $\alpha$.  The
average over $\vec{x}$ is symbolically implied in $\mathrm{Tr}$.
Then, an alternative definition of the Meissner screening mass is
immediately available from this thermodynamic potential as
\begin{equation}
 m_{M,\alpha\beta}^2 = \frac{1}{3}\sum_{i=1}^3
  \frac{\partial^2\Omega_A}{\partial A^i_\alpha \partial A^i_\beta}
  \Bigr|_{A=0}
  \sim 2\kappa^{ab}(T^\alpha)_{ca}(T^\beta)_{bd}\Delta^c\Delta^d \,,
\end{equation}
where $\Delta(x)$ is assumed to be spatial constant.  It might be
instructive to see how this expression works actually.  In the 2SC
phase for example, $\Delta^a\propto\delta^{a3}$ and
$(T^1)_{a3}=(T^2)_{a3}=(T^3)_{a3}=0$, and hence one can instantly
conclude that the $A_1$, $A_2$, and $A_3$ gluons are not Meissner
screened.  In the g2SC case the $A_8$ gluon is unstable, which can be
stated by the condition $\kappa^{33}<0$ because $m_{M,88}^2\sim
\kappa^{33}(\Delta^3)^2$.

  Now we shall assume the crystalline superconducting phase with
(\ref{eq:crystalline}).  The curvature of the thermodynamic potential
with respect to $\vec{q}$ is
\begin{equation}
 \frac{1}{3}\sum_{i=1}^3 \frac{\partial^2\Omega}{\partial q^i
  \partial q^i} \sim \kappa^{ab}\Delta^a\Delta^b.
\end{equation}
In the 2SC phase the instability condition for $\vec{q}$ to grow is
given by $\kappa^{33}<0$ again.  This argument works well only in the
(g)2SC case in which the $A_8$-instability can be identified as the
instability toward a crystalline superconducting phase equivalently.
Essentially the above is what has been articulated in
Refs.~\cite{Giannakis:2004pf,Huang:2005pv}.

  The generalization to the (g)CFL problem is possible with a simple
extension of the crystalline ansatz (\ref{eq:crystalline}).  As
suggested by Giannakis and Ren, we can consider a \textit{colored}
crystalline superconducting phase with the gap parameter taking a form
of
\begin{equation}
 \Delta(x) = |\Delta|e^{iT^\alpha \vec{q}^\alpha \cdot\vec{x}},
\label{eq:colored}
\end{equation}
then it is almost obvious that the curvature with respect to
$\vec{q}^\alpha$ is identical with the Meissner mass squared.  In this
sense, we can identify the instability regions in
Fig.~\ref{fig:phase_dg} with the colored crystalline phase with
nonvanishing $\vec{q}^\alpha$.

  What we should do next is now apparent; $\vec{q}^\alpha$ are the new
variational parameters to be determined so as to minimize the
thermodynamic potential.  This is a quite tough task, however.  The
number of the new variables is five corresponding to $A_{1,2}$,
$A_{4,5}$, $A_{6,7}$, $A_3$, and $A_8$.  To be worse, the rotational
symmetry is broken by the direction of $\vec{q}^\alpha$, and the
momentum angle integration cannot simplify in evaluating the
thermodynamic potential.  So, the numerical calculations are too
time-consuming to be done.  Moreover it is difficult to achieve an
enough accuracy to get reliable outputs in the multi-dimensional
numerical integration.  Maybe we have to abandon this apparent
strategy and instead need to invent a wiser simplification that would
not lose the important physics.

  One possibility would be to go to the higher orders in the
derivative expansion~(\ref{eq:expansion}).  The calculation would be
feasible because the expansion coefficients can be evaluated at
vanishing $\vec{q}^\alpha$ and the rotational symmetry is not broken
then.  Although we certainly have a chance to find the energy minimum
with some $\vec{q}^\alpha$, there is no guarantee that the next higher
order terms can be adequate to stabilize the potential.  In any case
this kind of calculation has yet to be performed.

  Finally, if I am excused to speak of my own opinion, all these
theoretical efforts will be tested in the lattice QCD simulation
someday when the sign problem at finite density will be solved and the
lattice spacing can be small enough to describe high density matter.
The phase diagram like Fig.~\ref{fig:phase_dg} is to be confirmed
then.
\vspace{3mm}

  This article is based on the talk given for the new talent sessions
at International School of Subnuclear Physics 43rd Course held at
Erice in Italy from Aug.~29 to Sep.~7 in 2005.  I thank all the
organizers for stimulating lectures and discussions at school.

%

\end{document}